\documentclass{article}

\usepackage{amsmath,amssymb,amsfonts}   
\usepackage{graphicx,bm}

\newcommand{\dis}{\displaystyle}
 
\renewcommand{\theequation}{\arabic{section}.\arabic{equation}}

\newcommand{\x}{{\bf x}}
\newcommand{\y}{{\bf y}}

\newcommand{\R}{\mathbb{R}}

\renewcommand{\u}{\widetilde{u}}

\newcommand{\Markov}[2]{\underset{#1}{\overset{#2}{\rightleftharpoons}}}

\newcommand{\A}{{\mathcal A}}

\newcommand{\calU}{{\mathcal U}}

\newcommand{\calA}{{\mathcal A}}
\newcommand{\calB}{{\mathcal B}}
\newcommand{\calJ}{{\mathcal J}}

\newcommand{\n}{\mathbf n}

\begin{document}

 \title{2D interfacial diffusion model of inhibitory synaptic receptor dynamics}


\author{ \em
P. C. Bressloff, \\ Department of Mathematics, 
University of Utah \\155 South 1400 East, Salt Lake City, UT 84112}

 \maketitle

\begin{abstract} 
The lateral diffusion and trapping of protein receptors within the postsynaptic membrane of a neuron plays a key role in determining the strength of synaptic connections and their regulation during learning and memory. In this paper we construct and analyze a 2D interfacial diffusion model of inhibitory synaptic receptor dynamics. The model involves three major components. First, the boundary of each synapse is treated as a semi-permeable interface due to the effects of cytoskeletal structures. Second, the effective diffusivity within a synapse is taken to be smaller than the extrasynaptic diffusivity due to the temporary binding to scaffold protein buffers within the synapse. Third, receptors from intracellular pools are inserted into the membrane extrasynaptically and internalized extrasynaptically and synaptically. We first solve the model equations for a single synapse in an unbounded domain and explore how the non-equilibrium steady-state number of synaptic receptors depends on model parameters. We then use matched asymptotic analysis to solve the corresponding problem of multiple synapses in a large, bounded domain. Finally, treating a synapse as a phase separated condensate of scaffold proteins, we describe how diffusion of individual scaffold proteins can also be modeled in terms of interfacial diffusion.
We thus establish interfacial diffusion as a general paradigm for exploring synaptic dynamics and plasticity.

\end{abstract}

\section{Introduction}

Recent advances in single particle tracking (SPT) and imaging techniques have established that the lateral diffusion and trapping of neurotransmitter receptors in

\noindent postsynaptic domains of a neuron plays a key role in mediating synaptic strength and plasticity \cite{Meier01,Borgdorff02,Dahan03,Choquet03,Bredt03,Groc04,Collinridge04,Triller05,Ashby06,Ehlers07,Groc08,Gerrow10,Henley11,Choquet13,Roth17,Choquet18,Triller19,Triller22}. Diffusion-trapping appears to be a general mechanism
for most types of excitatory and inhibitory neurotransmitter receptors. One well-studied example is the glycine receptor (GlyR), which mediates chloride-dependent synaptic inhibition
 in the postsynaptic membrane of the soma and initial portion of dendrites in spinal cord neurons. It has been observed experimentally that freely diffusing GlyRs are temporarily confined within post-synaptic densities (PSDs) that contain the scaffold protein gephyrin \cite{Meier01,Dahan03}. Surface receptors are also internalized via an active form of vesicular transport known as endocytosis, and then either recycled to the surface via exocytosis, or sorted for degradation  \cite{Ehlers00}. The majority of fast excitatory synapses in the central nervous system involve the neurotransmitter glutamate binding to $\alpha$-amino-3-hydroxy-5-methyl-4-isoxazole-propionic acid receptors (AMPARs). There are two major differences between glutamatergic synapses and inhibitory synapses. First, the former tend to be located along more distal regions of dendrites and are typically not found in the soma. Second, the PSD of an excitatory synapse is located within a dendritic spine, which is a small, sub-micrometer membranous
extrusion that protrudes from a dendrite \cite{Sorra00}. It is thought that the specialized geometry of the spine contributes to the trapping of AMPARs within the PSD, in addition to  interactions with scaffold proteins such as PSD-95 \cite{Borgdorff02,Groc04,Choquet18}. 

The quasi-one-dimensional (1D) geometry of a dendrite over length scales of the order of hundreds of microns has motivated a number of 1D diffusion-trapping models of AMPAR trafficking, in which the spines are represented as point sources or sinks at discrete locations along the dendritic cable, or as single compartments in a spatially discrete model \cite{Earnshaw06,Holcman06,Bressloff07,Earnshaw08,Thoumine12,Triesch18,Bressloff22}. These models typically assume that each synaptic compartment contains a fixed number of trapping sites or ``slots'' that temporarily bind receptors; these involve the various scaffold proteins within the cell membrane. The number of synaptic receptors is thus a dynamic steady-state that determines the strength of the synapse; activity-dependent changes in the strength of the synapse then correspond to shifts in the dynamical set-point. Note that on slower time scales, scaffold proteins and other synaptic components are also transported into and out of a synapse \cite{Salvatico15}. Hence, several experimental and modeling studies have analyzed the joint localization of gephyrin scaffold proteins and GlyRs at synapses \cite{Sekimoto09,Hasel11,Hasel15,Hakim20,Specht21}, showing how stable receptor-scaffold domains could arise dynamically. There is also recent evidence that the scaffold proteins form a phase separated biological condensate \cite{Zeng16,Bai21,Hosokawa21}.

In contrast to AMPAR trafficking in dendrites, the quasi-1D approximation is not appropriate for inhibitory synapses located in the somatic membrane, for example. In such cases, one has to treat the somatic membrane as a two-dimensional (2D) domain containing one or more synapses that act as transient traps. Moreover, one can no longer treat the synapses as point-like, since this leads to logarithmic singularities in the solution to the associated model equations. (On the other hand, in the case of a long, thin, spiny dendrite these singularities do not play a significant role, since the solutions of the 2D model are in excellent agreement with a reduced 1D model \cite{Bressloff08}.) Another advantage of working with a more realistic 2D model is that the size and shape of a synapse can be modeled explicitly.

In this paper we construct and analyze a 2D interfacial diffusion model of receptors by inhibitory synapses. Three major assumptions of our model are as follows. (i) The boundary of each synapse acts as an effective semi-permeable interface. This is motivated by the so-called partitioned fluid-mosaic model of the plasma membrane, in which confinement domains are formed by a fluctuating network of cytoskeletal fence proteins combined with transmembrane picket proteins that are anchored to the membrane skeleton \cite{Kusumi05}. The pickets thus act as posts along the membrane skeleton fence. (ii) The effective diffusivity within a synapse is much smaller than the extrasynaptic diffusivity due to molecular crowding and the temporary binding to scaffold proteins within the PSD \cite{Meier01,Dahan03,Groc08}. (iii) Receptors from intracellular pools are inserted into the membrane extrasynaptically and internalized both extrasynaptically and synaptically \cite{Ehlers00}. One major difference from previous 1D models is that we do not explicitly model receptor binding to slot proteins within a synapse. Instead, we encode the latter process in the synaptic diffusivity. That is, we consider a well-known biophysical mechanism for reducing the diffusion coefficient, in which scaffold proteins act as mobile buffers \cite{Keener09}. 

The structure of the paper is as follow. In section 2 we formulate our model for a single synapse in terms of a system of forward reaction-diffusion equations for the extrasynaptic and synaptic receptor concentrations. We assume that the particle flux across the semi-permeable interface is continuous, whereas there is a jump discontinuity in the concentration. One method for deriving the semi-permeable boundary conditions is to consider a thin membrane and to apply statistical thermodynamical principles \cite{Kedem58,Kedem62,Kargol96}. This leads to the so-called Kedem-Katchalsky equations, which also allow for discontinuities in the diffusivity and chemical potential across the interface; the latter introduces a directional bias. We include both features in the model. One consequence of including a directional basis is that the boundary conditions of the corresponding backward equations are not self-adjoint, which we establish in the appendix. In section 3, we explicitly solve the steady-state forward equations in the case of a circularly symmetric synapse in an unbounded domain. We then explore how the non-equilibrium steady-state number of synaptic receptors depends on the nature of the interface and various model parameters, including the rates of exocytosis and endocytosis, and the density of slot proteins. All of the latter have been identified as possible targets of stimulation protocols that induce synaptic modifications \cite{Bredt03,Henley11,Choquet18,Triller22}.  

In section 4 we extend the model to multiple synapses in a bounded domain $\Omega$. We use asymptotic perturbation methods to solve the reaction-diffusion equations in steady state along analogous lines to Refs. \cite{Ward93,Bressloff08,Coombs09,Ward15,Lindsay16,Bressloff21a,Bressloff22a}. This exploits the fact that the synapses are taken to be small compared to the size of the domain $|\Omega|$, that is, the ratio of each synaptic area relative to $|\Omega|$ is $O(\epsilon^2)$ with $0<\epsilon \ll 1$. The resulting asymptotic solution for the steady-state number of receptors in each synapse is non-perturbative in the sense that it sums over all logarithmic terms involving the small parameter $\nu=-1/\log \epsilon$. Expanding this solution in powers of $\nu$ we show that the leading order term in the expansion is independent of the other synapses and is consistent with single synapse result of section 2 in the small synapse limit. In addition, higher-order terms have a direct interpretation in terms of pairwise synaptic interactions mediated by bulk diffusion.
Finally, treating a synapse as a phase separated condensate of scaffold proteins, we describe in section 5 how diffusion of individual scaffold proteins can be modeled in terms of interfacial diffusion. We thus establish interfacial diffusion as a general paradigm for exploring synaptic receptor dynamics and plasticity.

\setcounter{equation}{0}
\section{Interfacial diffusion model}

\begin{figure}[t!]
\centering
\includegraphics[width=12cm]{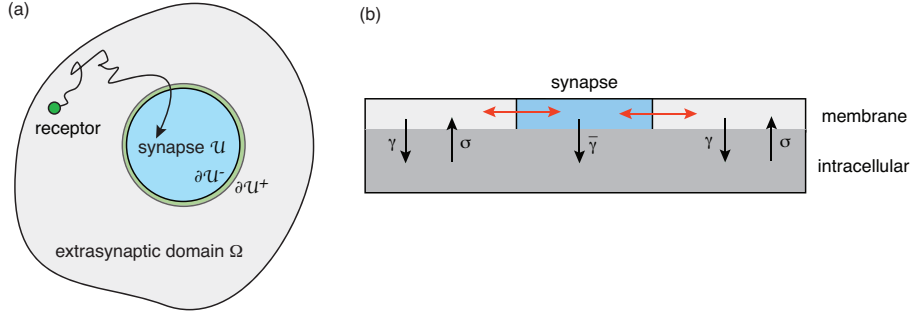} 
\caption{(a) Diffusion of receptors in a local region $\Omega$ of the post-synaptic membrane containing an inhibitory synapse $\calU$. The boundary $\partial \calU$ of the synapse is treated as a semi-permeable interface. (b) Receptors are cycled between the cell membrane and the interior of the cell via exocytosis and endocytosis. Exocytosis only occurs extransynaptically at a rate $\sigma$ whereas endocytosis occurs extransynaptically at a rate $\gamma$ and synaptically at a rate $\gamma_{-}$. Receptors also enter and exit the synapse via lateral diffusion.}
\label{fig1}
\end{figure}

Let $\Omega$ denote a local region of the plasma membrane that contains an inhibitory  synapse denoted by $\calU\subset \Omega$, see Fig. \ref{fig1}(a). For the moment, we assume that the exterior boundary $\partial \Omega$ is totally reflecting. Let $u(\x,t)$ denote the concentration (per unit area) of receptors with $u=u_+ $ for $\x\in \calU^c=\Omega\backslash \calU$ and $u=u_-$ for $\x \in \calU$. We represent the heterogeneous and crowded environment of the PSD as follows:
 \medskip
 
\noindent (i) The boundary $\partial \calU$ is treated as a semi-permeable interface with $\partial \calU^+ $ ($\partial \calU^-$) denoting the side approached from outside (inside) $\calU$. The particle flux across the interface is continuous but there is a jump discontinuity in the concentration.
\medskip

\noindent (ii) The diffusivity is assumed to be piecewise constant with $D(\x)=D$ for $\x\in \calU^c$ and $D(\x)=\overline{D}$ for $\x \in \calU$ with $\overline{D} <D$.
\medskip

\noindent (iii) Receptors from intracellular pools are inserted into the membrane extrasynaptically at a rate $\sigma$ and internalized both extrasynaptically and synaptically at the rates $\gamma$ and $\overline{\gamma}$, respectively, see Fig. \ref{fig1}(b).
\medskip

\noindent Under the above assumptions, the concentrations evolve according to the reaction-diffusion scheme
\begin{subequations} 
\label{master}
\begin{align}
&	\frac{\partial u_+(\x,t)}{\partial t} = D\nabla^2 u_+(\x,t)-\gamma u_+(\x,t)+\sigma, \, \x\in  \calU^c,\\
&	\frac{\partial u_-(\x,t)}{\partial t} = \overline{D} \nabla^2 u_-(\x,t)  -\overline{\gamma} u_-(\x,t), \ \x\in \calU,  \\
&	D\nabla u_+(\x,t)\cdot \n=0,\ \x \in \partial \Omega , \\
&	D\nabla u_+(\x^+,t)\cdot \n_0 = \overline{D}\nabla u_-(\x^-,t)\cdot \n_0\equiv J(\x,t) \\
	&J(\x,t)=\kappa [(1-\alpha )u_+(\x^+,t)- \alpha u_-(\x^-,t)] ,\quad x^{\pm} \in \partial \calU^{\pm}.
\end{align}
	\end{subequations}
Here $\n$ is the outward unit normal at a point on $\partial \Omega$, $\n_0$ is the outward unit normal at a point on $\partial \calU$, $J(\x,t)$ is the continuous inward flux across the point $\x\in \partial \calU$, $\kappa$ is the permeability of the interface $\partial \calU$, and $\alpha\in [0,1]$ specifies a directional bias with $\alpha =1/2$ the unbiased case. Equations (\ref{master}d,e) are one version of the well-known Kedem-Katchalsky (KK) equations \cite{Kedem58,Kedem62,Kargol96}. Finally, we impose the initial conditions $u_+(\x,0)=0$ and $ u_-(\x,0)=0$, $j=1,\ldots,N$.
Finally, the number of receptors within the synapse is
\begin{equation}
\label{rk}
r(t)=\int_{\calU}u_-(\x,t)d\x.
\end{equation}

\subsection{Diffusivity and buffering}

Suppose that synaptic receptors can reversibly bind to scaffold proteins according to the reaction scheme
\[R+S \Markov{k_-}{k_+} RS.\]
Let $w(\x,t)$ and $w^*(\x,t)$ denote the concentration of free and bound scaffold proteins respectively.
The reaction diffusion equations within the synapse become
\begin{subequations} 
\label{RDbuff}
\begin{align}
&	\frac{\partial u_-(\x,t)}{\partial t} = D_{\rm syn}\nabla^2 u_-(\x,t) +k_-w^*(\x,t)-k_+u_-(\x,t)w(\x,t)  -\gamma_{\rm syn} u_-(\x,t),   \\
&\frac{\partial w(\x,t)}{\partial t} =D_{w}\nabla^2 w(\x,t)  +k_-w^*(\x,t)-k_+u_-(\x,t)w(\x,t),   \\
&\frac{\partial w^*(\x,t)}{\partial t} = D_{w}\nabla^2 w^*(\x,t)  -k_-w^*(\x,t)	+ k_+u_-(\x,t)w(\x,t).\end{align}
	\end{subequations}
We allow for the possibility that the diffusivity $D_{\rm syn}$ and rate of endocytosis $\gamma_{\rm syn}$ of unbound receptors within a synapse differ from their extrasynaptic counterparts. We also assume that unbound and bound scaffold proteins have the same diffusivity $D_{w}$. Adding equations (\ref{RDbuff}b,c) shows that
	\begin{equation}
	\frac{\partial (w(\x,t)+w^*(\x,t))}{\partial t} =D_{w}\nabla^2 (w(\x,t)+w^*(\x,t)).
	\end{equation}
	Assuming that $w+w^*$ is initially uniform, then it remains uniform for all time and we can set $w(\x,t)+w^*(\x,t)=\overline{w}$, where $\overline{w}$ is the total amount of scaffold protein. If the binding/unbinding reaction rates $k_{\pm}$ are sufficiently fast then we can assume that $v$ and $w$ are in quasi-equilibrium:
	\begin{equation}
	k_-(\overline{w}-w(\x,t))=k_+u_-(\x,t)w(\x,t),
	\end{equation}
	which implies that
\begin{equation}
\label{wv}
w(\x,t)=\frac{K_{\rm eq} \overline{w}}{K_{\rm eq}+u_-(\x,t)},\quad K_{\rm eq}=\frac{k_-}{k_+}.
\end{equation}

Subtracting equations (\ref{RDbuff}a,b) yields
\begin{align}
\frac{\partial [u_-(\x,t)-w(\x,t)}{\partial t} &= D_{\rm syn}\nabla^2 u_-(\x,t)-D_{w} \nabla^2 w(\x,t)  -\gamma_{\rm syn} u_-(\x,t).
\end{align}
Substituting for $w$ using equation (\ref{wv}) and using the chain-rule for differentiation gives
\begin{align}
&\left (1+\frac{K_{\rm eq} \overline{w}}{(K_{\rm eq}+u_-(\x,t))^2}\right )\frac{\partial u_-(\x,t)}{\partial t}=D_{\rm syn}\nabla^2 u_-(\x,t)\\
&\quad +D_{w}\nabla \cdot \frac{K_{\rm eq} \overline{w}}{(K_{\rm eq}+u_-(\x,t))^2}\nabla u_-(\x,t)-\gamma_{\rm syn} u_-(\x,t).\nonumber
\end{align}
We now make two further simplifications. First we take $D_{w}\ll D_{\rm syn}$ so that the second term on the right-hand side can be dropped. Second, we assume that $u_-(\x,t)\ll K_{\rm eq}$. It follows that 
\begin{equation}
\frac{\partial u_-(\x,t)}{\partial t}=\frac{D_{\rm syn}}{1+\overline{w}/K_{\rm eq}} \nabla^2 u_-(\x,t)-\frac{\gamma_{\rm syn}}{1+\overline{w}/K_{\rm eq}} u_-(\x,t)
\end{equation}
and, hence,
\begin{equation}
\label{Dbar}
\overline{D}=\frac{D_{\rm syn}}{W_0},\quad \overline{\gamma}=\frac{\gamma_{\rm syn}}{W_0},\quad W_0=1+\overline{w}/K_{\rm eq}. 
\end{equation}
Note that
\begin{equation}
\label{Dgam}
\frac{\overline{D}}{\overline{\gamma}}=\frac{D_{\rm syn}}{\gamma_{\rm syn}}.
\end{equation}
Moreover, a synapse with a larger ``synaptic weight'' $\overline{w}$ has a smaller diffusivity and a lower rate of endocytosis, both of which serve to enhance the number of receptors within the synapse.

\subsection{Steady-state equations}

The steady state distribution of synaptic receptors can be determined by setting all time derivatives to zero so that equations (\ref{master}) become
\begin{subequations}
\label{masterLT}
\begin{align}
 &D\nabla^2u_+(\x)- \gamma u_+(x) =-\sigma,  \, \x\in   \calU^c,\\
&  \overline{D}\nabla^2u_-(\x)- \overline{\gamma} u_-(\x)=0, \ \x\in   \calU,\\
&D\nabla u_+(\x) \cdot \n=0,\ \x \in \partial \Omega,\\
&	D\nabla u_+(\x^+)\cdot \n_0 = \overline{D}\nabla u_-(\x^-)\cdot \n_0=J(\x),\\
	&J(\x)=\kappa [(1-\alpha) u_+(\x^+) -\alpha u_-(\x^-)] ,\quad \x^{\pm} \in \partial \calU^{\pm}.
 \end{align}
\end{subequations}
Integrating equation (\ref{masterLT}a) with respect to $\x\in  \calU^c$ and using the divergence theorem yields 
\begin{align}
&D\int_{\partial \Omega} \nabla  u_+(\x)\cdot \n d\ell_{\x}-D \int_{\partial \calU} \nabla  u_+(\x)\cdot \n_0 d\ell_{\x}+\sigma|\Omega|=  0.\nonumber \\
\end{align}
From the boundary conditions (\ref{masterLT}c,d), we then have
\begin{align}
& -\int_{\partial \calU}J(\x)d\ell_{\x}+\sigma|\Omega|= \gamma\int_{\calU^c} u_+(\x)d\x.
\end{align}
Finally, integrating equation (\ref{masterLT}b) with respect to $\x\in \calU$ shows that
\begin{equation}
\int_{\partial \calU}J(\x)d\ell_{\x}= \overline{\gamma} r^*,
\end{equation}
where $r^*=\int_{\calU}u_-(\x)d\x$ is the steady-state number of synaptic receptors.
Hence, we obtain the steady-state conservation equation
\begin{align}
&\sigma |\Omega|=\gamma\int_{ \calU^c} u_+(\x)d\x +\overline{\gamma} r^*.
\label{con1}
\end{align}
The left-hand side is the total constant flux entering the domain $\calU^c$ due to exocytosis, which is balanced on the right-hand side by the total steady-state rate of receptor internalization due to endocytosis.

\subsection{First passage times and the backward diffusion equation}

The macroscopic model of receptor interfacial diffusion given by equations (\ref{master}) is useful for exploring how the non-equilibrium steady-state number of receptors within the synapse depends on the number of slot proteins and other synaptic parameters. However, given advances in SPT experiments, there is also considerable interest in the statistics of individual receptor trajectories, including first passage time statistics. For the sake of illustration, consider the MFPT for a surface receptor to be internalized via endocytosis. In order to calculate the MFPT and related quantities, it is necessary to consider a single-particle version of equations (\ref{master}). That is, $u=u(\x,t|\x_0)$ is interpreted as the probability density that a labeled receptor is at position $\x\in \Omega$ at time $t$, given that it started at $\x_0$, and $\sigma=0$. As previously, we set $u=u_+(\x,t|\x_0)$ for $\x \in \calU^c$ and $u=u_-(\x,t|\x_0)$ for $\x \in \calU$.

Let $Q(\x_0,t)$ denote the survival probability, which is defined according to
\begin{equation}
Q(\x_0,t)=\int_{\calU^c}u_+(\x,t|\x_0)d\x+\int_{\calU}u_-(\x,t|\x_0)d\x.
\end{equation}
Differentiating both sides with respect to $t$ and using equations (\ref{master}a,b) shows that
\begin{align}
 \frac{\partial Q(\x_0,t)}{\partial t}&=D\int_{\calU^c}\nabla^2u_+(\x,t|\x_0)d\x +\overline{D}\int_{\calU} \nabla^2 u_-(\x,t|\x_0)d\x\nonumber \\
 &\quad -\gamma\int_{\calU^c}u_+(\x,t|\x_0)d\x -\overline{\gamma}\int_{\calU}  u_-(\x,t|\x_0)d\x\nonumber \\
 &=D  \int_{\partial \Omega}\nabla u_+\cdot \n d\sigma-D  \int_{\partial \calU^+}\nabla u_+\cdot \n_0 d\sigma+\overline{D} \int_{\partial \calU^-}\nabla u_-\cdot \n_0 d\sigma\nonumber \\
 &\quad -\gamma\int_{\calU^c}u_+(\x,t|\x_0)d\x -\overline{\gamma}\int_{\calU}  u_-(\x,t|\x_0)d\x .
\label{Q2}
\end{align}
It follows from the boundary conditions (\ref{master}c-e) that all boundary fluxes disappear so that
\begin{equation}
\frac{\partial Q(\x_0,t)}{\partial t}=-\gamma\int_{\calU^c}u_+(\x,t|\x_0)d\x -\overline{\gamma}\int_{\calU}  u_-(\x,t|\x_0)d\x\equiv \calJ(\x_0,t),
\end{equation}
where $\calJ(\x_0,t)$ is the total endocytotic flux.
Laplace transforming equation (\ref{Q2}) and imposing the initial condition $Q(\x_0,0)=1$ gives
\begin{equation}
\label{QL}
s\widetilde{Q}(\x_0,s)-1=-  \widetilde{\calJ}(\x_0,s).
\end{equation}
Since $\Omega$ is bounded, and assuming that $\kappa>0$, the particle is eventually absorbed with probability one, which means that
$\lim_{t\rightarrow \infty}Q(\x_0,t)=\lim_{s \rightarrow 0}s\widetilde{Q}(\x_0,s) =0$. Hence, $\widetilde{\calJ}(\x_0,0)=1$. Finally, using the fact that $-\partial_tQ(\x_0,t)$ is the FPT density, we have
\begin{equation}
\label{Tuncon}
T(\x_0)=-\int_0^{\infty}t\partial_tQ(\x_0,t) dt=\int_0^{\infty}Q(\x_0,t)dt=\widetilde{Q}(\x_0,0) = -\left . \frac{\partial}{\partial s} \widetilde{\calJ}(\x_0,s)\right |_{s=0}.
\end{equation}

It follows from equation (\ref{Tuncon}) that there are two complementary methods for calculating $T(\x_0)$. The first is to solve the forward equation (\ref{master}) in Laplace space and use this to evaluate the first derivative of the flux $\calJ(\x_0,s)$ in the limit $s\rightarrow 0$. The second is to construct the corresponding backward equation for the survival probability and MFPT with respect to $\x_0$. As far as we are aware, the backward equation in the case of the general asymmetric boundary conditions (\ref{master}c) is not well-known so we derive it explicitly in the appendix. We simply summarize the result here. For all $\x \in \Omega$, let $u(\x,t|\y)=v_+(\x,t|\y)$ if $\y \in  \calU^c$ and $u(\x,t|\y)=v_-(\x,t|\y)$  if $\y \in  \calU$. The backward equations are then
\begin{subequations} 
\label{masterb}
\begin{align}
	\frac{\partial v_+(\x,t|\y)}{\partial t} &= D\nabla_{\y}^2 v_+(\x,t|\y)-\gamma v_+(\x,t|\y) , \ \y\in \calU^c,\\
	\frac{\partial v_-(\x,t|\y)}{\partial t} &= \overline{D}\nabla_{\y}^2 v_-(\x,t|\y)  -\overline{\gamma}v_-(\x,t|\y),\ \y\in \calU,
	\end{align}
	 with the adjoint semi-permeable boundary conditions
\begin{align}
\alpha D\nabla_{\y} v_+(\x,t|\y^+)\cdot \n_0  &= (1-\alpha) \overline{D}\nabla_{\y} v_-(\x,t|\y^-)\cdot \n_0\equiv K(\x,t|\y), \\
K(\x,t|\y)&= \kappa \alpha(1-\alpha) [v_+(\x,t|\y^+)-v_-(\x,t|\y^-)] ,\ \y\in \partial \calU,
\end{align}
and the exterior boundary condition
\begin{equation}
 \nabla v_+(\x,t|\y) \cdot \n=0,\ \y \in \partial \Omega .
\end{equation}
\end{subequations}
It can be seen that for all $\alpha\neq 1/2$, the semi-permeable boundary conditions are not self-adjoint. 

We can now write down the corresponding backward equation for the MFPT. First, note that  integrating equations (\ref{masterb}) with respect to $\x\in \Omega$, setting $\y=\x_0$ and taking all derivatives with respect to $\x_0$ shows that the survival probability also satisfies the backward diffusion equation. Laplace transforming the latter gives
\begin{subequations} 
\label{masterQ}
\begin{align}
	s\widetilde{Q}(\x_0,s) -1 &= D\nabla^2 \widetilde{Q}(\x_0,s)-\gamma\widetilde{Q}(\x_0,s), \ \x_0\in \calU^c, \ \nabla \widetilde{Q}(\x_0,s)\cdot \n =0,\ \x_0 \in \partial \Omega, \\
	s\widetilde{Q}(\x_0,s) -1 &= \overline{D}\nabla^2 \widetilde{Q}(\x_0,s) -\overline{\gamma}\widetilde{Q}(\x_0,s)  ,\ \x_0\in \calU,
	\end{align}
	 with the adjoint semi-permeable boundary conditions
\begin{align}
\alpha D\nabla  \widetilde{Q}(\x_0^+,s) \cdot \n_0  &= (1-\alpha) \overline{D}\nabla  \widetilde{Q}(\x_0^-,s) \cdot \n_0 \nonumber \\
&= \kappa \alpha(1-\alpha) [\widetilde{Q}(\x_0^+,s) -\widetilde{Q}(\x_0^-,s) ] ,\ \x_0\in \partial \calU .
\end{align}
\end{subequations}
Finally, taking the limit $s\rightarrow 0$ yields
\begin{subequations}
\label{BVPT}
\begin{align}
D\nabla^2 T(\x_0)-\gamma T(\x_0)&=-1, \  \x_0\in \calU^c, \quad \nabla T(\x_0)\cdot \n=0,\ \x_0\in \partial \Omega\\
\overline{D}\nabla^2 T(\x_0)-\overline{\gamma}T(\x_0) &=-1, \  \x\in \calU, \\
\alpha D\nabla T(\x_0^+)\cdot \n_0  &= (1-\alpha) \overline{D}\nabla  T(\x_0^-) \cdot \n_0\nonumber \\
&= \kappa \alpha(1-\alpha) [T(\x_0^+)-T(\x_0)^-] ,\ \x_0\in \partial \calU.
\end{align}
\end{subequations}

\setcounter{equation}{0}
\section{Example with circular symmetry}

In order to illustrate the above theory, suppose that $\Omega=\R^2$ and $\calU$ is a disc of radius $\rho_1$, $\calU =\{\x\in \R^2\,|\, 0 <  |\x| <\rho_1\}$.  We will calculate the steady state number of receptors $r^*$. The analysis of the MFPT proceeds along analogous lines. Introducing polar coordinates with $\rho =|\x|$, the steady-state equations (\ref{masterLT}) become
\begin{subequations}
\label{circ}
\begin{align}
 &D\frac{\partial^2u_+(\rho)}{\partial \rho^2} + \frac{D}{\rho}\frac{\partial u_+(\rho)}{\partial \rho} - \gamma u_+(\rho) =-\sigma,  \, \rho_1<\rho ,\\
&  \overline{D}\frac{\partial^2u_-(\rho)}{\partial \rho^2} + \frac{\overline{D}}{\rho}\frac{\partial u_-(\rho)}{\partial \rho} - \overline{\gamma}u_-(\rho) =0, \ 0<\rho<\rho_1,\\
&D\partial_{\rho}u_+(\rho_2)  =0,\\
&	D\partial_{\rho} u_+(\rho_1^+) = \overline{D}\partial_{\rho} u_-(\rho_1^-)=\kappa [(1-\alpha) u_+(\rho_1^+) -\alpha u_-(\rho_1^-)].
 \end{align}
\end{subequations}
 Equations of the form (\ref{circ}) can be solved in terms of modified Bessel functions. The general solution is given by
 \begin{subequations}
 \label{qir}
  \begin{align}
     u_+(\rho) &= A_+  K_0(\beta \rho) +\frac{\sigma}{\gamma}, \ \rho_1<\rho<\rho_2,\\
      u_-(\rho) &= A_-  I_0(\beta_{\rm syn} \rho),\quad 0<\rho <\rho_1,
\end{align}
\end{subequations}
with
\begin{equation}
\beta=\sqrt{\frac{\gamma}{D}},\quad \beta_{\rm syn}=\sqrt{\frac{\overline{\gamma}}{\overline{D}}}=\sqrt{\frac{\gamma_{\rm syn}}{D_{\rm syn}}}. 
\end{equation}
We have used equation (\ref{Dgam}).
In addition, $I_{0}$ and $K_{0}$ are zeroth order modified Bessel functions of the first and second kind, respectively. 
For simplicity, we set $D=D_{\rm syn}$ and introduce the dimensionless variables $\ell_j=\beta \rho_j$, $j=1,2$.

The unknown coefficients $A_{\pm}$ are determined from the boundary conditions (\ref{circ}c,d):
\begin{subequations}
\begin{align}
\sqrt{\gamma_{-}  \overline{D}} A_- I_0'(\beta_{\rm syn} \rho_1)&=\sqrt{\gamma D}  A_+   K'_0(\beta \rho_1)),\\
  &=\kappa\left \{(1-\alpha)\left [A_+  K_0(\beta \rho_1)+\frac{\sigma}{\gamma}\right ]-\alpha A_-I_0(\beta_{\rm syn} \rho_1)\right \}.
 \end{align}
\end{subequations}
Rearranging these equations yields
\begin{subequations}
\begin{align}
A_-&=-\frac{ \sqrt{\gamma D}}{ \sqrt{\gamma_{-}  \overline{D}}}\frac{K'_0(\beta \rho_1)}{I_0'(\beta_{\rm syn} \rho_1)}A_+,\\
\frac{\sigma}{\gamma}&=\left (\frac{\sqrt{\gamma_{-}  \overline{D}} }{\kappa(1-\alpha)} I_0'(\beta_{\rm syn} \rho_1)+\frac{\alpha}{1-\alpha} I_0(\beta_{\rm syn} \rho_1)\right )A_- -K_0(\beta \rho_1)  A_+\nonumber \\
&=-\left \{\frac{ \sqrt{\gamma D}}{ \sqrt{\gamma_{-}  \overline{D}}}\left (\frac{\sqrt{\gamma_{-}  \overline{D}}  }{\kappa(1-\alpha)} +\frac{\alpha}{1-\alpha} \frac{I_0(\beta_{\rm syn} \rho_1)}{I_0'(\beta_{\rm syn} \rho_1)}\right ) K'_0(\beta \rho_1) +K_0(\beta \rho_1)\right \} A_+ .
\end{align} 
\end{subequations}
Note the Bessel identities $I'_0(x)=I_1(x)$ and $K_0'(x)=-K_1(x)$.

 \begin{figure}[b!]
\centering
\includegraphics[width=12cm]{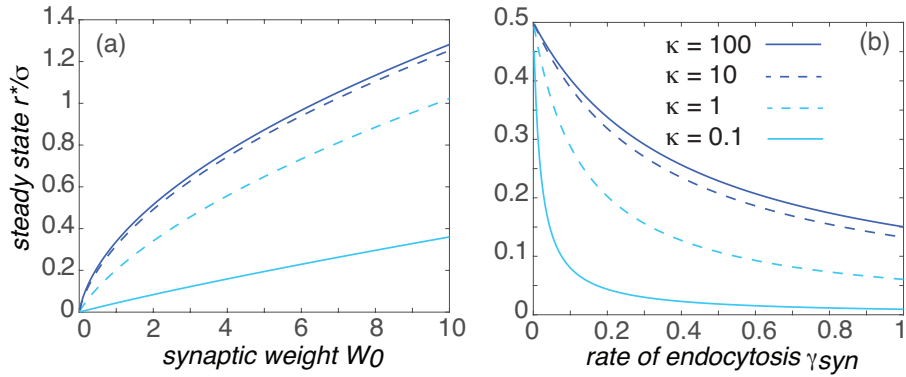} 
\caption{Plot of the (normalized) steady-state number of synaptic receptors $r^*/\sigma$ versus (a) the weight $W_0$ and (b) the rate of endocytosis $\gamma_{\rm syn}$ for various values of $\kappa$. Default parameter values are $W_0=1$ and $\gamma=D=\gamma_{\rm syn}=D_{\rm syn}=1$.}
\label{fig2}
\end{figure}

\begin{figure}[t!]
\centering
\includegraphics[width=12cm]{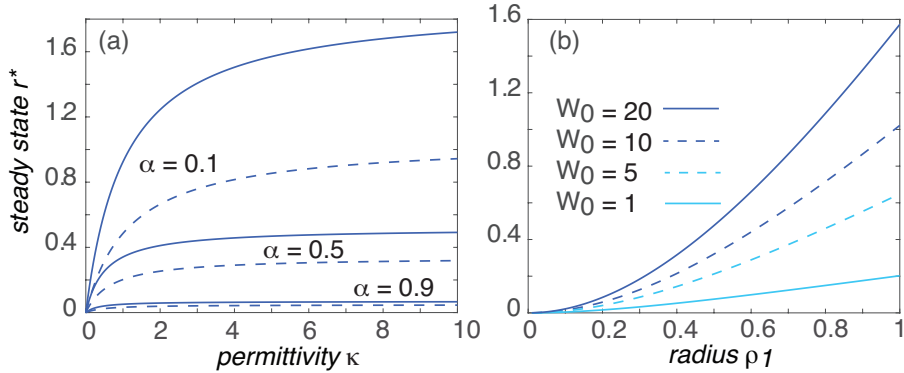} 
\caption{(a) Plot of the (normalized) steady-state number of synaptic receptors $r^*$ versus the permittivity $\kappa$ for $W_0=2$ (solid curves) and $W_0=1$ (dashed curves) and various values of $\alpha$. We also take $ \rho_1=1$. (b) Plot of $r^*$ versus the radius $\rho_1$ of the synapse for various $W_0$ and $\alpha=0.5$. Other parameter values are $\gamma=\gamma_{\rm syn}=D=D_{\rm syn}=1$.}
\label{fig3}
\end{figure}

From equations (\ref{rk}) and (\ref{qir}b), the steady-state number of synaptic receptors is
\begin{align}
r^*&= \int_{\calU}u_-(\x)d\x=   2\pi\int_0^{\rho_1} u_-(\rho)\rho d\rho =2\pi A_-\int_0^{\rho_1}I_0\left(\beta_{\rm syn}\rho \right)\rho d\rho.
\label{ber}
\end{align}
From the Bessel function identity $\frac{d}{dx}[xI_1(x)]=xI_0(x)$,
we have
\begin{align*}
\int_0^{\rho_1} I_0(\beta_{\rm syn} \rho)\rho d\rho&=\beta_{\rm syn}^{-2} \int_0^{\beta_{\rm syn} \rho_1}x I_0(x)dx
=\frac{1}{\beta_{\rm syn}^{2}} \int_0^{\beta_{\rm syn}\rho_1}\frac{d}{dx} [x I_1(x)]dx\\
&=\frac{\rho_1}{\beta_{\rm syn}}I_1(\beta_{\rm syn} \rho_1).
\end{align*}
Hence,
\begin{align}
r^*&=\frac{ \sqrt{\gamma D}}{ \sqrt{\gamma_{-}  \overline{D}}}\frac{\dis  K_1(\beta \rho_1)}{\dis  \frac{ \sqrt{\gamma D}}{ \sqrt{\gamma_{-}  \overline{D}}}\left (\frac{\sqrt{\gamma_{-}  \overline{D}}}{\kappa(1-\alpha)} +\frac{\alpha}{1-\alpha} \frac{I_0(\beta_{\rm syn} \rho_1)}{I_0'(\beta_{\rm syn} \rho_1)}\right )K_1(\beta \rho_1)+K_0(\beta \rho_1)}\frac{2\pi \sigma \rho_1}{\gamma   \beta_{\rm syn}}\nonumber \\
&=\frac{2\pi D\sigma}{\gamma\gamma_{\rm syn}}\frac{\dis  W_0K_1(\beta \rho_1)\beta \rho_1}{\dis \sqrt{\gamma D}\left [ \frac{1}{\kappa(1-\alpha)} +\frac{\alpha }{1-\alpha}\frac{\sqrt{W_0}}{\sqrt{\gamma_{\rm syn}D_{\rm syn}}}\Theta( \beta_{\rm syn} \rho_1) \right ]K_1(\beta \rho_1)+ K_0(\beta \rho_1)} ,
\label{rrs}
\end{align}
where $\Theta(z)={I_0(z)}/{I_1(z)}$.
We have used equation (\ref{Dbar}).  A number of remarks are in order. \medskip

 \noindent (a) Experimentally determined values of the rate of internalization and extrasynaptic diffusivity are $\gamma  \sim 10^{-3}-10^{-2} s^{-1}$ and $D\sim 0.1 \mu m^2s^{-1} $, respectively \cite{Ehlers07}. This implies that the fundamental length scale $\sqrt{D/\gamma}\sim 1-10\mu $m. Since the size of a typical synaptic domain is a micron, we will take $\beta \rho_1\sim 0.1-1$.  We fix the units of time by setting $\gamma=D=1$ and assume that $D_{\rm syn}=1$. In dimensionless units, we can write
 \begin{align}
\label{rss3}
r^*&= \frac{\sigma}{\gamma_{\rm syn}}\frac{\dis   W_0K_1( \rho_1) \rho_1}{\dis \left [ \frac{1}{\kappa(1-\alpha)} +\frac{\alpha }{1-\alpha} \sqrt{\frac{W_0}{\gamma_{\rm syn}}}\Theta( \rho_1 \sqrt{\gamma_{\rm syn}}) \right ]K_1(\rho_1)+ K_0( \rho_1)} .\end{align}
\medskip
 
 \noindent (b) 
It can be seen that there are at least three distinct mechanisms for increasing $r^*$: (i) increasing the rate of extrasynaptic exocytosis $\sigma$; (ii) increasing $W_0$ by increasing the density of slot proteins $\overline{w}$;  (iii) reducing the rate of synaptic endocytosis $\gamma_{\rm syn}$. All three mechanisms have been observed to play a role in long term synaptic potentiation \cite{Bredt03,Henley11,Choquet18,Triller22}. Note, that $r^*$ is a linear function of $\sigma$ but a nonlinear function of $W_0$ and $\gamma_{\rm syn}$, see Fig. \ref{fig2}.
 \medskip
 
 \noindent (c) The steady-state $r^*$ also depends on the semi-permeable membrane parameters. In the limits $\kappa\rightarrow 0$ (impermeable boundary $\partial \calU$) or $\alpha \rightarrow 1$ (zero flux into the synapse) the steady-state number of receptors is zero. In Fig. \ref{fig3}(a) we plot the steady-state $r^*$ as a function of $\kappa$ for various $\alpha$ and $W_0$. As expected, the number of synaptic receptors is a monotonically increasing function of $\kappa$ and a monotonically decreasing value of $\alpha$. To what extent variations in $\kappa$ and $\alpha$ play a role in synaptic plasticity is less clear. Finally, we find that $r^*$ is a monotonically increasing function of the synaptic radius $\rho_1$, see in Fig. \ref{fig3}(b), which is also consistent with experimental studies.

\setcounter{equation}{0}
\section{Multiple synapses in a bounded domain}

Now suppose that there are $N$ synapses $\calU_j\subset \Omega$, $j=1,\ldots,N$. The multi-synaptic version of the interfacial diffusion model takes the form
\begin{subequations} 
\label{RD}
\begin{align}
&	\frac{\partial u(\x,t)}{\partial t} = D\nabla^2 u(\x,t)-\gamma u(\x,t)+\sigma, \, \x\in \Omega\backslash \calU_a,\quad \calU_a=\bigcup_{j=1}^N\calU_j,\\
&	\frac{\partial u_j(\x,t)}{\partial t} = D_j\nabla^2 u_j(\x,t)  -\gamma_j u_j(\x,t), \ \x\in \calU_j,  \\
&	D\nabla u(\x,t)\cdot \n=0,\ \x \in \partial \Omega , \\
&	D\nabla u(\x^+,t)\cdot \n_j = D_j\nabla u_j(\x^-,t)\cdot \n_j\equiv J_j(\x,t), \\
	&J_j(\x,t)=\kappa_j [(1-\alpha_j )u(\x^+,t)- \alpha_ju_j(\x^-,t)] ,\quad \x^{\pm} \in \partial \calU_j^{\pm},
\end{align}
	\end{subequations}
	so that the number of receptors in the $j$th synapse is given by
	\begin{equation}
	\label{rj}
	r_j(t)=\int_{\calU_j}u_j(\x,t)d\x.
	\end{equation}
We have added the subscript $j$ to each of the synaptic parameters $D_j,\gamma_j,\kappa_j,\alpha_j,\overline{w}_j$ with
\begin{align}
D_j&=\frac{D_{\rm syn}}{1+\overline{w}_j/K_{\rm eq}},\quad \gamma_j=\frac{\gamma_{\rm syn}}{1+\overline{w}_j/K_{\rm eq}}.
\label{Dj}
\end{align}
The analysis of equations (\ref{RD}) is considerably more involved, even when $\Omega=\R^2$ 
Therefore, in order to make the problem analytically tractable, we take the area of each synapse to be much smaller than $|\Omega|$, that is, $|\calU_j|\sim \epsilon^2 |\Omega|$. For concreteness, each synapse is  a disc of radius $\rho_j=\epsilon \ell_j$ centered about a point $\x_j \in \Omega$: $\calU_j=\{\x \in \calU, \ |\x-\x_j|\leq \epsilon \ell_j\}$, $j=1,\ldots,N$. In addition, the synapses are assumed to be well separated with $|\x_i-\x_j|=O(1)$, $j\neq i$, and $\mbox{dist}(\x_j,\partial \Omega)=O(1)$ for all $j=1,\ldots,N$.. Under these various assumptions, the resulting system of equations can be solved by extending asymptotic and Green's function methods previously developed for the so-called 2D narrow capture problem \cite{Ward93,Bressloff08,Coombs09,Ward15,Lindsay16,Bressloff21a,Bressloff22a}.

\subsection{Scaling}

The basic asymptotic method involves constructing an inner or local solution valid in the interior and an $O(\epsilon)$ exterior neighborhood of each synapse, and then matching to an outer or global solution that is valid away from each neighborhood. Before proceeding, however, we need to make certain assumptions about how various model parameters scale with $\epsilon$. Intuitively speaking, in order to have $O(1)-O(100)$ receptors $r_j$ in each synapse $j=1,\ldots,N$, we require large synaptic and scaffold protein concentrations within the $O(\epsilon^2)$-sized PSD domains, that is, $u_j=O(1/\epsilon^2)$ and $\overline{w}_j=O(1/\epsilon^2)$. However, this would lead to a large outflow through each semi-permeable interface unless $\alpha_j=O(\epsilon^2)$ in equation (\ref{RD}e). The latter can be justified using the buffering model of sect. 2, since the semi-permeable interface would be impermeable or weakly permeable with respect to receptors bound to scaffold proteins. Next, the flux continuity condition (\ref{RD}d) across the interface suggests that we should take $D_j=O(\epsilon^2)$. This also follows from equation (\ref{Dj}). On the other hand, we assume $\gamma_{\rm syn}=O(1/\epsilon^2)$ so that $\gamma_j=O(1)$. Finally, since the circumference of each synapse $\calU_j$ is $O(\epsilon)$ we take the permeability $\kappa_j=O(1/\epsilon)$ so that the total flux into each synapse is $O(1)$. We summarize the various parameter scalings as follows:
\begin{align}
\label{scale}
\overline{w}_j\rightarrow \epsilon^{-2} \overline{w}_j,\ \alpha_j\rightarrow \epsilon^2\alpha_j, \ D_j\rightarrow \epsilon^2 D_j,\ \kappa_j\rightarrow \kappa_j/\epsilon,\quad \gamma_{\rm syn}\rightarrow \gamma_{\rm syn}/\epsilon^2.
\end{align}
In particular, note from equation (\ref{Dj}) that the rescaled synaptic diffusivity satisfies
\begin{align}
D_j&=\frac{1}{\epsilon^2}\frac{D_{\rm syn}}{1+\epsilon^{-2}\overline{w}_jk/K_{\rm eq}}\approx \frac{K_{\rm eq}D_{\rm syn}}{\overline{w}_j}.
\label{Dj2}
\end{align}
Under these various parameter rescalings, the steady-state version of equations (\ref{RD}) become
\begin{subequations}
\label{masterLTs}
\begin{align}
 &D\nabla^2u(\x)- \gamma {u}(\x) =-\sigma,  \, \x\in \Omega\backslash \calU_a,\\
& \epsilon^2 D_j\nabla^2u_j(x)-\gamma_j u_j(\x)=0, \ \x\in   \calU_j,\\
&D\nabla u(\x) \cdot \n=0,\ \x \in \partial \Omega,\\
&	D\nabla u(\x^+)\cdot \n_j = \epsilon^2D_j\nabla u_j(\x^-)\cdot \n_j  \equiv \epsilon^{-1} {J}_j(\x),\\
	&{J}_j(\x)=\kappa_j [(1-\epsilon^2 \alpha_j) u(\x^+) -\epsilon^2 \alpha_ju_j(\x^-)] ,\quad \x^{\pm} \in \partial \calU_j^{\pm}.
 \end{align}
\end{subequations}

\subsection{Inner solution}

The inner solution near the $j$-th synapse is constructed by introducing the stretched local variable ${\mathbf y} =
\varepsilon^{-1}(\x-\x_j)$ and setting
\begin{equation}
U(\y)=u(\x_j+\varepsilon \y),\quad U_j(\y)=\epsilon^2 u_j(\x_j+\varepsilon \y).
\end{equation}
The inner solutions $U,U_j$ then satisfy (on dropping $O(\epsilon)$ terms)
\begin{subequations}
\label{inner}
\begin{align}
&D \nabla^2_{\y}U(\y)= 0,\ |\y| > \ell_j,\\ &D_j\nabla^2_{\y}U_j (\y)- \gamma_j U_j(\y)= 0 , \   |\y| < \ell_j ,\\
	&D\nabla U(\y^+)\cdot \n_j = D_j\nabla U_j(\y^-)\cdot \n_j =\kappa_j [U(\y^+)-\alpha_jU_j(\y^-)],\quad   |\y^{\pm} |=\ell_j^{\pm} .
\end{align}
\end{subequations}
Using polar coordinates with $|\y|=\rho$, the solution can be written as
\begin{subequations}
\label{UV}
\begin{align}
    &U =  \Phi_j^+ +C_j \log\rho/\ell_j ,\quad \ell_j \leq \rho < \infty, \\
    &{U}_j=\Phi_j ^-\frac{I_0\left(\beta_{\rm syn}\rho\right)}{I_0(\beta_{\rm syn} \ell_j)},\quad 0 \leq \rho \leq \ell_j .
\end{align}
\end{subequations}
The coefficients $\Phi_j^{\pm}$ can be expressed in terms of $C_j$ by imposing equations (\ref{inner}c):
\begin{align}
\label{UV2}
   \frac{C_jD}{D_j} &= {\mathcal F}(\beta_{\rm syn}\ell_j)\Phi_j ^-, \quad
\frac{C_jD}{\kappa_j\ell_j} =  \Phi_j^+- \alpha_j\Phi_j^-,\quad {\mathcal F}(x)=\frac{xI_1\left(x\right)}{I_0(x)}.
\end{align}
Rearranging shows that $\Phi_j^{\pm}=\Psi_j^{\pm}C_j$ with
\begin{align}
\Psi_j^- &=\frac{D}{D_j{\mathcal F}(\beta_{\rm syn} \ell_j)},\quad
\Psi_j^+ = \left [\frac{D}{\kappa_j\ell_j}+ \frac{\alpha_jD }{D_j{\mathcal F}(\beta_{\rm syn} \ell_j)}\right ] .
\label{Psip}
\end{align}
In order to determine the remaining $N$ coefficients $C_j$ we have to match the far-field behavior of the inner solution $U$ with the outer solution.

\subsection{Matching with the outer solution}

The outer solution is constructed by shrinking each synapse to a single point and imposing a corresponding singularity condition that is obtained by matching with the inner solution. The outer equation is given by
\begin{align}
\label{outer}
	 D\nabla^2 u(\x) - \gamma u(\x)= -\sigma
		\end{align}
for $\x\in \Omega'\equiv \Omega\backslash \{\x_1,\ldots,\x_N\},$ together with the boundary condition $D\nabla u \cdot \n=0, \ \x\in \partial \Omega$. The corresponding singularity conditions are
\begin{align}
\label{goo}
    \u \sim     C_j \left[\Psi_j ^++ \log|{\bf x} - {\bf x}_j|/\ell_j-\log \epsilon \right]
\end{align}
for ${\bf x} \to {\bf x}_j$. A common feature of strongly localized perturbations in 2D domains \cite{Ward93} is the appearance of the small parameter
\begin{equation}
\nu=-\frac{1}{\log \epsilon }.
\end{equation}
In order to eliminate the $1/\nu$ term in the singularity condition (\ref{goo}), we rescale the unknown coefficients $C_j$ by setting $C_j=\nu \A_j(\nu)$.
 It is well-known that $\nu \rightarrow 0$ more slowly than $\epsilon\rightarrow 0$. Hence, if one is interested in obtaining $O(1)$ accuracy with respect to an $\epsilon$ expansion, then it is necessary to sum over the logarithmic terms non-perturbatively \cite{Ward93}.

The first step is to introduce the Green's function of the modified Helmholtz equation according to
\begin{subequations}
\begin{align}
\label{GMH}
	&-\delta(\x-\y) = D\nabla^2 G(\x,\y) - \gamma G(\x,\y), \ \x\in \Omega,\\
	&0 = \nabla G(\x,\y)\cdot \n ,\ \x \in \partial \Omega,\quad \int_{\Omega} G(\x,\y)d\x=\frac{1}{\gamma} . 
	\end{align}
	\end{subequations}
Note that $G$ can be decomposed as 
\begin{align}
    G(\x,\y) = -\frac{\log{|\x - \y|}}{2\pi D} + R(\x, \y),
\end{align}
where $R$ is the non-singular part of the Green's function. It follows that the solution of equation (\ref{outer}) can be written in the form 
\begin{align}
    u(\x) &\sim  \frac{\sigma}{ \gamma}     - 2\pi \nu D\sum_{j=1}^N \A_j(\nu)G(\x,\x_j).
    \label{outu}
\end{align}
We have $N$ unknown coefficients $\A_j(\nu)$, which are determined by matching the inner and outer solutions:
\begin{align}
   &  \frac{\sigma}{ \gamma} 
    = \A_j \left[1 + \nu \Psi_j^++ 2\pi  D \nu R(\x_j,\x_j )\right]+ 2\pi  D \nu \sum_{i \neq j}A_i G({\bf x}_i, \x_j) .
  \label{match}
\end{align}
Let us rewrite equation (\ref{match}) as a matrix equation:
\begin{align}
\label{matrix}
    \left[{\bf I} + \nu\left({\bm \Psi}^+ + 2\pi  D{\bf G} ^{\top}\right)\right]{\bf a} = \frac{\sigma}{\gamma}{\bf I},
\end{align}
where ${\bf I}$ is the $N \times N$ identity matrix, 
\begin{align}
{\bf a}=(\A_1(\nu ), \ldots, \A_N(\nu))^{\top},\quad {\bm \Psi}^+=\mbox{diag}(\Psi_1^+,\ldots,\Psi^+_N), 
\end{align}
and ${\bf G}$ is an $N \times N$ matrix with entries
\begin{equation}
G_{ij}=G(\x_i,\x_j),\ i\neq j,\quad G_{jj}=R(\x_j,\x_j).
\end{equation}
Inverting equation (\ref{matrix}) yields in component form
\begin{align}
\label{naj}
    \A_j(\nu) 
    &= \sum_{i=1}^N\left({\bf I}  + \nu {\bf M} \right)^{-1}_{ji}\overline{G}_i,\quad j = 1, \ldots, N,
    \end{align}
 with
  \begin{equation}
  \label{M}
 M_{ji}=\Psi_j^+\delta_{i,j} +2\pi  D G_{ij}.
    \end{equation}
 This is a non-perturbative solution that sums over all logarithmic terms involving factors of $\nu$, along analogous lines to \cite{Ward93}. It is $O(1)$ with respect to a corresponding $\epsilon$ expansion.   

In summary, given the solution to equation (\ref{naj}), we have the following approximations for the bulk and synaptic receptor concentrations:
\begin{subequations}
\label{fUV}
    \begin{align}
    u(\x) &\sim \frac{\sigma}{ \gamma}- 2\pi \nu D\sum_{j=1}^n A_j(\nu)G(\x,\x_j),\ \x\in \Omega\backslash \{\x_1,\ldots,\x_N\},\\
   u_j(\x)&\sim\frac{1}{\epsilon^2} \frac{\nu D\A_j(\nu)}{D_j}\frac{I_0\left(\beta_{\rm syn}|\x-\x_j|/\epsilon\right)}{\beta_{\rm syn}\ell_jI_1(\beta_{\rm syn}\ell_j)},\quad 0 \leq |\x-\x_j| \leq \epsilon \ell_j.\end{align}
\end{subequations}
From equations (\ref{rj}) and (\ref{fUV}b), the steady-state number of receptors in the $j$-th synapse is
\begin{align}
r_j^*&= \int_{\calU_j}u_j(\x)d\x\sim   \frac{2\pi \nu D\A_j(\nu)}{D_j}\int_0^{\ell_j}\frac{I_0\left(\beta_{\rm syn}\rho \right)}{\beta_{\rm syn} \ell_jI_1(\beta_{\rm syn}\ell_j)}\rho d\rho\nonumber \\
&=\frac{2\pi \nu D\A_j(\nu)}{\beta_{\rm syn}^2D_j}=\frac{\overline{w}_j}{K_{\rm eq} }\frac{2\pi \nu  D}{\gamma_{\rm syn}}{\A}_j(\nu).
\label{rstar}
\end{align}
We evaluated the integral along identical lines to the analysis of equation (\ref{ber}) and used equation (\ref{Dj2}).
 Setting
 \begin{equation}
 \A_j(\nu)=\A_j^{(0)}-\nu \A_j^{(1)}+\nu^2 \A_j^{(2)}-\ldots
 \end{equation}
and Taylor expanding the right-hand side of equation (\ref{naj}) with respect to $\nu$ gives
\begin{subequations}
  \begin{align}
\A_j^{(0)}&= \frac{\sigma}{\gamma},\quad
\A_j^{(1)}(\nu)=  \frac{\sigma}{\gamma}\left\{\Psi_j^+(0)  + 2\pi D\sum_{i=1}^NG_{ij}(0)  \right \},\\
\A_j^{(2)}(\nu)&= \frac{\sigma}{\gamma}\bigg\{\Psi_j^+(0)\Psi_j^+(0) 
  +2\pi D \sum_{l=1}^N (\Psi_j^+(0)+\Psi_l^+(0))G_{lj}(0) \nonumber \\
&\quad +(2\pi D)^2\sum_{k,l=1}^NG_{kj}(0)G_{lk}(0)\bigg \}.
 \end{align}
 \end{subequations}
 The $\nu$-expansion of the coefficient $\A_j(\nu)$ can be represented diagrammatically as shown in Fig. \ref{fig5} to $O(\nu^3)$. A number of results emerge from our analysis:
\medskip

\begin{figure}[t!]
\centering
\includegraphics[width=12cm]{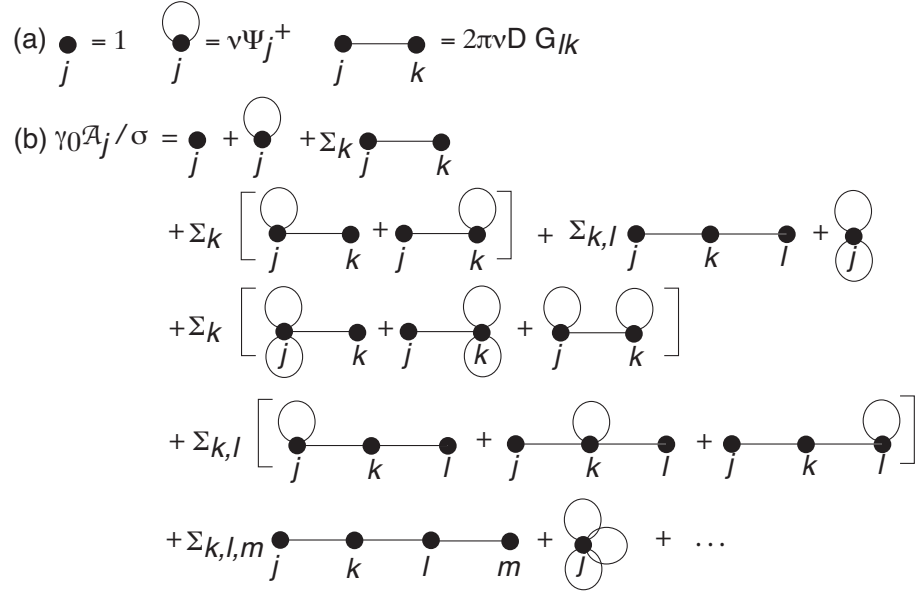} 
\caption{Diagrammatic representation of the perturbation expansion of $\A_j(\nu)$ given by equation (\ref{naj}). (a) Basic elements used in the diagrams. Each loop associated with a synapse $j$ represents insertion of a factor $\nu \Psi_j^+$, whereas the line connecting a pair of synapses $(j,k)$ represents the propagator $2\pi D\nu G_{jk}$. (b) Set of diagrams up to $O(\nu^3)$. }
\label{fig5}
\end{figure}

\noindent (i) The steady-state number of receptors in the $j$-th synapse is independent of the other synapses to leading order in $\nu$: 
\begin{align}
\label{rss2D}
r_j^* &\sim \frac{\sigma}{\gamma} \frac{\overline{w}_j}{K_{\rm eq} }\frac{2\pi \nu  D}{\gamma_{\rm syn}} +O(\nu^2). 
\end{align}
In particular, the leading order term is independent of the semi-permeable interface parameters $\kappa_j,\alpha_j$.
As a quick check, we show that equation (\ref{rss2D}) is consistent with our exact result (\ref{rrs}) for a single synapse in $\R^2$. Setting $\rho_1=\epsilon \ell_1$ and introducing the scalings (\ref{scale}), we have
\begin{align}
\label{rss}
r^*
&=\frac{2\pi D\sigma \epsilon^2 }{\gamma\gamma_{\rm syn}}\\
&\times \frac{\dis  (1+\overline{w}/\epsilon^2K_{\rm eq})K_1(\beta \epsilon \ell_1)\beta \epsilon \ell_1}{\dis \frac{\sqrt{\gamma D}}{(1-\epsilon ^2\alpha)}\left [ \frac{\epsilon}{\kappa} + \epsilon^2\alpha \frac{\epsilon \sqrt{1+\overline{w}/\epsilon^2K_{\rm eq}}}{\sqrt{\gamma_{\rm syn}D_{\rm syn}}}\Theta( \beta_{\rm syn} \ell_1) \right ]K_1(\beta \epsilon \ell_1)+ K_0(\beta \epsilon \ell_1)} .\nonumber
\end{align}
Using the leading order expansions
\[ I_0(z)\sim 1,\quad I_1(z)\sim z,\quad K_0(x)\sim -\ln z, \quad K_1(z)\sim \frac{1}{z}\]
for $z\rightarrow 0$, we see that
\begin{align}
r^*
&\approx \frac{2\pi D\sigma  }{\gamma\gamma_{\rm syn}}\frac{\overline{w}}{K_{\rm eq}}\frac{1}{\dis \sqrt{\gamma D}\left [ \frac{\epsilon}{\kappa} + \epsilon^2\frac{\alpha \overline{w}}{K_{\rm eq}} \frac{\Theta( \beta_{\rm syn} \ell_1) }{\sqrt{\gamma_{\rm syn}D_{\rm syn}}}\right ]\frac{1}{\beta \epsilon \ell_1}-\log(\beta \epsilon \ell_1)} \approx \frac{\sigma}{\gamma} \frac{\overline{w}}{K_{\rm eq} }\frac{2\pi \nu  D}{\gamma_{\rm syn}}.
\end{align}

\medskip

\noindent (ii) The full non-perturbative solution for $\A_j(\nu)$ involves the coefficients $\Psi_j^+$, which depend on $\kappa_j$, $\alpha_j$, $\ell_j$ and $D_j$ via equation (\ref{Psip}). (The insertion of these coefficients is represented by the single-site loops in Fig. \ref{fig5}.) This means that the simple multiplicative dependence on the density of slot proteins and various global parameters breaks down at higher orders in $\nu$. Moreover, decreasing the permeability $\kappa_j$ or increasing the outward bias $\alpha_j$ leads to an increase in $\Psi_j^+$ and a concomitant reduction in the number of synaptic receptors $r_j$.
\medskip

\noindent (iii) The full non-perturbative solution for $\A_j(\nu)$ also introduces synaptic coupling that is mediated by bulk diffusion via the Green's function $G(\x_i,\x_j)$. The synaptic interactions are represented by the propagators in Fig. \ref{fig5}. An analogous result was previously obtained in a 1D diffusion-trapping model of dendritic receptor trafficking \cite{Bressloff22a}, although a different scaling regime was considered due to the absence of logarithmic singularities. As shown in the 1D model, the interaction terms could provide a substrate for heterosynaptic forms of plasticity, particularly as $\nu$ is a slowly decreasing function of $\epsilon$.

\setcounter{equation}{0}
\section{Interfacial diffusion, scaffold proteins and liquid-liquid phase separation}

A number of recent experimental studies have suggested receptor-dependent liquid-liquid phase separation as a possible mechanism of PSD formation and maintenance \cite{Zeng16,Bai21,Hosokawa21}. Such a mechanism was also proposed in an earlier theoretical paper \cite{Sekimoto09}, where scaffold-receptor aggregation was taken to be the result of spontaneous phase separation between a dilute phase of extrasynaptic scaffold proteins and a dense phase of synaptic scaffold proteins, which also depended on the colocalization of the cognate receptors. All populations were assumed to be in thermodynamic equilibrium, at least on time scales faster than the timescale of protein degradation.

In the case of equilibrium phase-separated liquids, it has recently been shown that individual molecular trajectories can be described in terms of a Langevin equation with drift in an effective potential that has a steep gradient at the phase boundary \cite{Julicher21}. (Analogous to advances in the single-particle tracking of synaptic receptors, it is now possible to image individual molecules crossing condensate boundaries \cite{Mora21}. Hence, developing stochastic models of single-molecule trajectories is of considerable current interest.) In Ref. \cite{Julicher21} the Langevin equation was solved in a 1D domain with a phase boundary at $x=0$. Motivated by the application to PSDs, suppose that at equilibrium the dense phase is localized in a circular droplet $\calU$ of radius $\rho_1$ and the dilute phase is in the bounded domain $\calU_c=\Omega\backslash \calU$, see Fig. \ref{fig1}. Let $p_{-}(\x,t|\x_0)$, $\x\in \calU$, denote the probability density that a solute molecule is in the dense phase and let $p_+(\x,t|\x_0)$, $\x \in \calU^c$, be the corresponding density in the dilute phase.  Following Ref. \cite{Julicher21}, the single-particle diffusion equations in the sharp interface limit are of the form
\begin{subequations} 
\label{masterp}
\begin{align}
&	\frac{\partial p_+(\x,t|\x_0)}{\partial t} = D\nabla^2 p_+(\x,t|\x_0)\, \x\in  \calU^c,\\
&	\frac{\partial p_-(\x,t|\x_0)}{\partial t} = \overline{D} \nabla^2 p_-(\x,t|\x_0)  \ \x\in \calU,  \\
&	D\nabla p_+(\x,t|\x_0)\cdot \n=0,\ \x \in \partial \Omega , \\
&	D\nabla p_+(\x^+,t|\x_0)\cdot \n_0 = \overline{D}\nabla p_-(\x^-,t|\x_0)\cdot \n_0\equiv J(\x,t) ,\\
	&p_-(\x^-,t|\x_0)= \Gamma p_+(\x^+,t)|\x_0] ,\quad \x^{\pm} \in \partial \calU^{\pm},\quad \Gamma = \frac{\phi_-}{\phi_+},
\end{align}
	\end{subequations}
	where $\phi_-$ ($\phi_+$) is the equilibrium volume fraction of solute in the dense (dilute) phase, $D$ is the solute diffusivity in the dilate phase, and $\overline{D}$ is the solute diffusivity in the dense phase. For example, the diffusivities may depend on the volume fraction according to the relations $D=D_0(1-\phi_+)$ and $\overline{D}=D_0(1-\phi_-)$ \cite{Julicher21}. We now observe that, mathematically speaking, the interfacial boundary conditions (\ref{masterp}d,e) are a special limit of the semi-permeable boundary conditions (\ref{master}d,e). In particular, the former are obtained from the latter by taking the limit $\kappa \rightarrow \infty$ and setting $\Gamma=(1-\alpha)/\alpha$. This suggests that interfacial diffusion could also be used to model the stochastic dynamics of an individual scaffold protein, assuming that the PSD  can be treated as a scaffold protein condensate in quasi-equilibrium. 
	
One non-trivial issue concerns the presence of multiple synaptic condensates.
In the classical theory of liquid-liquid phase separation via spinodal decomposition, there is a rapid demixing from one thermodynamic phase to two coexisting phases due to the fact that there is essentially no thermodynamic barrier to nucleation of the two phases. In early stages of phase separation, solute molecules form microscopic solute-rich domains dispersed throughout the liquid. These droplets then rapidly grow and coalesce to form macroscopic droplets. Finally, in the late stages of phase separation, a diffusion-mediated form of coarsening known as Ostwald ripening ultimately results in the complete separation of the dilute phase from a single large droplet in the dense phase \cite{Weber19}.
A major difference between classical physical condensates and biological condensates is that the latter are often driven away from equilibrium by various energy-consuming processes. A number of theoretical studies have shown that the active regulation of liquid-liquid phase separation by adenosine triphosphate (ATP)-driven enzymatic reactions can suppress Ostwald ripening, resulting in the coexistence of multiple large droplets \cite{Zwicker15,Weber17a,Wurtz18,Lee19,Weber19,Bressloff20}. An alternative mechanism for coexistence has recently been found in an experimental study of P granules in the nematode worm {\em Caenorhabditis elegans} \cite{Folkmann21}. (P granules are biological condensates that are rich in RNA specific to the germline, and are located in the region of the cytoplasm near the cell nucleus.) In contrast to active phase separation, the P granules are stabilized by protein clusters that adsorb to the condensate interface, without interfering with the diffusive exchange of solute molecules across the interface. This is analogous to the role of so-called pickering agents in stabilizing inorganic emulsions \cite{Ramsden04,Pickering07}, which have a wide range of applications in the drug and food industries. 
As far as we are aware, the stabilization mechanism for PSD condensates is not yet known. If it turns out to be a passive mechanism, then one could develop a multi-synaptic version of equations (\ref{masterp}) and use the matched asymptotic methods of section 4 to analyze single particle diffusion across the phase boundaries. (Such methods have previously been used to study the active suppression of Ostwald ripening \cite{Bressloff20}.) However, in order to develop a more complete model it will be necessary to understand the possible role of interactions between receptors and scaffold proteins during phase separation, and how this in turn affects receptor trafficking.

\section{Discussion}

The key idea of this paper is that interfacial diffusion provides a general paradigm for exploring synaptic receptor dynamics. In particular, we showed that the non-equilibrium steady state number of receptors in a synapse depends on the nature of the interface and the modified diffusive environment within a synapse due to the presence of scaffold proteins that stabilize the receptors. All of these are possible targets of stimulus protocols that induce various forms of synaptic plasticity. One potential limitation of our interfacial diffusion model is that it treats a synapse as spatially homogeneous. However, it is known that the PSDs of inhibitory synapses are partitioned into nanodomains, which are distinct regions of clustered molecules such as the scaffold proteins gephyrin \cite{Crosby19,Yang21,Triller22}. These nanodomains are thought to align bound postsynaptic receptors with corresponding presynaptic active domains. In order to incorporate such details into the interfacial model, we would need to treat the diffusion coefficient within a synapse as spatially dependent or uses some form of homogenization scheme. Interestingly, the heterogeneity of the PSD can be incorporated into phase separated models of the PSD, by taking into account the effects of signaling molecules within the condensate \cite{Zeng16,Bai21,Hosokawa21}.

Finally, note that it is also possible to consider a more general probabilistic framework for modeling diffusion across a semi-permeable interface using so-called snapping out Brownian motion (BM)\cite{Lejay16,Bressloff22I}. The basic idea is to sew together successive rounds of reflected BM that are restricted to either $\calU$ or $\calU^c$, see Fig. \ref{fig1}. Each round is killed when its Brownian local time (time spent in a neighborhood of either $\calU^+$ or $\calU^-$) exceeds a random threshold. A new round is then immediately started in one of the two domains that is selected probabilistically. If the threshold distribution is an exponential with rate $2\kappa$, then we recover the standard semi-permeable boundary condition used in equations (\ref{master}). However, taking the threshold distribution to be non-Markovian results in a time-dependent permeability that may be heavy-tailed \cite{Bressloff22I}. Given the complex molecular environment of the cell membrane and PSDs, it is not unreasonable to expect that some form of anomalous behavior occurs.

\setcounter{equation}{0}
\renewcommand{\theequation}{A.\arabic{equation}}

\section*{Appendix}

The classical method for deriving the backward diffusion equation is to note that the full solution $u(\x,t|\x_0)$ satisfies the Chapman-Kolmogorov equation \cite{Gardiner09,Bressloff21}
\begin{equation}
\label{ChapK}
u(\x,t|\x_0)=\int_{\Omega} u(\x,t-\tau|\y)u(\y,\tau|x_0)d\y, \quad 0\leq \tau <\leq t.
\end{equation}
That is, the probability of being at $\x$ at time $t$, given the initial position $\x_0$, is the sum of the probabilities of each possible path from $\x_0$ to $\x$. We have also imposed time translation invariance.
Since the left-hand side of equation (\ref{ChapK}) is independent of the intermediate time $\tau$, it follows that
\begin{align}
0&=\int_{\Omega}\partial_{\tau}u(\x,t-\tau|\y)u(\y,\tau|\x_0)d\y +\int_{\Omega}u(\x,t-\tau|\y)\partial_{\tau}u(\y,\tau|\x_0)d\y .
\end{align}
Since $u(\y,\tau|\x_0)$ satisfies the forward BVP equation, $\partial_{\tau}[u(\y,\tau |\x_0)]$ can be replaced by terms involving derivatives 
with respect to $\y$:
\begin{align*}
\int_{\Omega}u(\x,t-\tau|\y)\partial_{\tau}u(\y,\tau|\x_0)d\y &=\int_{\calU^c} v_+(\x,t-\tau|\y)[D\nabla^2_{\y}u_+(\y,\tau|\x_0)-\gamma u_+(\y,\tau|\x_0)]d\y\\
& +  \int_{ \calU} v_-(\x,t-\tau|\y)[\overline{D}\nabla^2_{\y}u_-(\y,\tau|\x_0)-\overline{\gamma}u_-(\y,\tau|\x_0)]d\y.
\end{align*}
(We have introduced that notation that for all $\x \in \Omega$, $u(\x,t-\tau|\y)=v_+(\x,t-\tau|\y)$ if $\y \in  \calU^c$ and $u(\x,t-\tau|\y)=v_-(\x,t-\tau|\y)$  if $\y \in  \calU$.)
Integrating by parts twice using Green's identities shows that
\begin{align*}
&\int_{\calU_c} v_+(\x,t-\tau|\y)\nabla^2_{\y}u_+(\y,\tau|\x_0)d\y\\
&=\int_{\calU_c} \nabla_{\y}^2 v_+(\x,t-\tau|\y) u_+(\y,\tau|\x_0) d\y- \int_{\partial \Omega} u_+(\y,\tau|\x_0) \nabla_{\y} v_+(\x,t-\tau|\y)\cdot \n d\y\\
&\quad + \int_{\partial \Omega} v_+(\x,t-\tau|\y)\nabla_{\y}u_+(\y,\tau|\x_0)\cdot \n \, d\y-   \int_{\partial \calU^+} v_+(\x,t-\tau|\y) \nabla_{\y} u_+(\y,\tau|\x_0) \cdot \n_k d\y \\
&\quad +   \int_{\partial \calU^+}  u_+(\y,\tau|\x_0) \nabla_{\y} v_+(\x,t-\tau|\y)\cdot \n_kd\y.
\end{align*}
Similarly,
\begin{align*}
\int_{ \calU} v_-(\x,t-\tau|\y) \nabla^2_{\y}u_-(\y,\tau|\x_0)&=\int_{\calU} \nabla_{\y}^2 v_-(\x,t-\tau|\y) u_-(\y,\tau|\x_0) d\y\\
&\quad +\int_{\partial \calU^-} v_-(\x,t-\tau|\y^-) \nabla_{\y} u_-(\y,\tau|\x_0) \cdot \n_0 d\y \\
&\quad -  \int_{\partial \calU^-}  u_-(\y,\tau|\x_0) \nabla_{\y} v_-(\x,t-\tau|\y)\cdot \n_0d\y.
\end{align*}

We see from the above equations that the various boundary terms cancel if we impose the adjoint boundary conditions $ \nabla_{\y} v_+(\x,t-\tau|\y)\cdot \n =0$ and
\begin{align*}
&D u_+(\y^+,\tau|\x_0) \nabla_{\y} v_+(\x,t-\tau|\y^+)\cdot \n_0-Dv_+(\x,t-\tau|\y^+) \nabla_{\y} u_+(\y^+,\tau|\x_0) \cdot \n_0 \\
&=\overline{D}u_-(\y^-,\tau|\x_0) \nabla_{\y} v_-(\x,t-\tau|\y^-)\cdot \n_0-\overline{D}v_-(\x,t-\tau|\y^-) \nabla_{\y} u_-(\y^-,\tau|\x_0) \cdot \n_0
\end{align*}
for all $\y^{\pm}\in \partial \calU^{\pm}$. Setting 
\begin{equation}
J^{\dagger}(\x,t|\y^+) = D  \nabla_{\y} v_+(\x,t|\y^+)\cdot \n_0,\quad J^{\dagger}(\x,t|\y^-) =\overline{D}  \nabla_{\y} v_-(\x,t-\tau|\y^-)\cdot \n_0
\label{Jdag}
\end{equation}
gives
\begin{align*}
& u_+(\y^+,\tau|\x_0) J^{\dagger}(\x,t-\tau|\y^+) -v_+(\x,t-\tau|\y^+) J(\y,\tau|\x_0)  \\
&=  u_-(\y^-,\tau|\x_0) J^{\dagger}(\x,t-\tau|\y^-) - v_-(\x,t-\tau|\y^-) J(\y,\tau|\x_0) ,\ \y^{\pm}\in \partial \calU^{\pm}.
\end{align*}
Rearranging the last equation and imposing the forward semi-permeable boundary conditions (\ref{master}d,e) yields
\begin{align}
& u_+(\y^+,\tau|\x_0) J^{\dagger}(\x,t-\tau|\y^+) - u_-(\y^-,\tau|\x_0) J^{\dagger}(\x,t-\tau|\y^-) \nonumber \\
&=  [v_+(\x,t-\tau|\y^+) - v_-(\x,t-\tau|\y^-)] J(\y,\tau|\x_0) \\
&=[v_+(\x,t-\tau|\y^+) - v_-(\x,t-\tau|\y^-)] \kappa[(1-\alpha)u_+(\y^+,\tau|\x_0)-\alpha u_-(\y^-,\tau|\x_0)]\nonumber 
\end{align}
for $\y^{\pm}\in \partial \calU^{\pm}$. We thus have the adjoint equations
\begin{subequations}
\label{Jdag2}
\begin{align}
J^{\dagger}(\x,t|\y^+) &= \kappa(1-\alpha) [v_+(\x,t|\y^+)-v_-(\x,t|\y^-)]\\
J^{\dagger}(\x,t|\y^-) &= \kappa \alpha[v_+(\x,t|\y^+)-v_-(\x,t|\y^-)] .
\end{align}
\end{subequations}
Given that all boundary terms vanish if equations (\ref{Jdag}) and (\ref{Jdag2}) hold, we finally obtain the backward equations (\ref{masterb}).

\end{document}